\documentstyle[times,pramana,epsf,floats]{ias}
\begin{document}
\mark{{Canonical quantum gravity and consistent discretizations}{Gambini and Pullin}}
\title{Canonical quantum gravity and consistent discretizations}

\author{Rodolfo Gambini}
\address{Instituto de F\'{\i}sica, Facultad de Ciencias, Igua esq. Matajo,
Montevideo, Uruguay}
\author{Jorge Pullin}
\address{Department of Physics and Astronomy, Louisiana State University,
Baton Rouge, LA 70803 USA}

\pacs{4.65}
\abstract{
This is a summary of the talk presented by JP at ICGC2004. It covered 
some developments in canonical quantum gravity occurred since 
ICGC2000, emphasizing the recently introduced consistent discretizations
of general relativity.}

\maketitle
\section{Introduction}

Canonical quantum gravity is the attempt to apply the rules of quantum
mechanics in their most traditional canonical form to general
relativity.  A first question that arises concerning this field of
endeavor is: why?  After all, attempts to canonically quantize general
relativity were done in the 1960's, notably by DeWitt following
earlier work by Dirac and Bergmann, and significant obstacles were
found. Moreover, it is not expected that general relativity will be
the ultimate theory of nature, and some proposals for such theories,
like string theory, appear to offer advantages towards quantization.
Our personal point of view on this issue is the following: there are
good reasons why we have had so much trouble quantizing general
relativity.  It is the first theory we try to quantize that is
invariant under diffeomorphisms (therefore it is a ``topological
quantum field theory'') that has local field-theoretic degrees of
freedom. Some experience with topological theories was gained in the
80's and 90's, but the examples considered only had a finite number of
degrees of freedom, i.e. they were really mechanical systems disguised
as field theories. Now, it is  expected that  any ultimate theory of
nature will be invariant under diffeomorphisms and will have local degrees
of freedom.  Therefore it is natural to first address these
difficulties in the simplest possible scenario of a theory with some
claim to describe certain aspects of nature, as general
relativity. What is learned will be of future use to tackle more
elaborate theories, as for instance string theory.  Therefore we do not
see any conflict between having a group of people working on canonical
quantum gravity while other groups study string theories or other
theories that attempt to unify all interactions.

The plan of this article is present a very brief and quick introduction
to results since ICGC2000. Due to limitations in the time format of the
talk we will only be able to cover a fraction of what has been achieved,
so we concentrate largely on issues which we had something to do with.
In particular we will not be able to cover attractive work on the
definition of semiclassical states by Ashtekar et al. \cite{shadow},
Thiemann et al. \cite{Thiemannsemi} and Varadarajan \cite{madhavan}.
I will not be able to discuss the work of Samuel on understanding the
real Ashtekar variables four dimensionally \cite{Samuel}, nor the
recent work on quasi-normal modes and the quantum of area \cite{Olaf}.
Developments on quantum cosmology with the loop formulation are occurring
rapidly and will be covered in the talk by Martin Bojowald. I will not
cover attempts to define the path integral formulation (spin-foams) nor
Thiemann's ``Phoenix project'' proposal \cite{phoenix}.

\section{Recent history}

I start with a brief recent history of the field to put the current
situation in perspective. Traditionally, canonical formulations of
general relativity considered as canonical variables the metric on a
spatial slice $q^{ab}$ and a canonically conjugate momentum closely
related to the extrinsic curvature of the slice. In 1985 Ashtekar
\cite{Ashtekar} noted that one could use an alternative set of
canonical variables.  The variables consisted of a set of (densitized)
triads $\tilde{E}^a_i$ and as canonically conjugated momentum a
(complex) $SO(3)$ connection $A_a^i$. The spatial metric (doubly
densitized) can be easily reconstructed as 
$\tilde{\tilde{q}}^{ab}=\tilde{E}^a_i \tilde{E}^b_i$. In terms of
these variables, the canonical theory has the usual constraints of canonical
gravity and an additional Gauss law. This is appealing since it
suggests that the phase space of general relativity can be viewed as a
subspace of the phase space of Yang--Mills theories. The usual
constraints of canonical gravity adopt rather simple expressions in
terms of these variables, they are given by polynomial expressions.
This initially raised hopes that the quantization could proceed in a
better way with this formulation than with the usual one.

The new variables change immediately the perspective on quantization.
In particular, the most natural quantum representation to consider is
to pick wavefunctions that are functionals of the connection
$\Psi[A]$.  The Gauss law constraint (promoted to a quantum operator)
implies that these functions have to be gauge invariant functions of
the connection.  In 1988 Rovelli and Smolin \cite{RoSm} noted that one
could use as a basis of functions of the connection the traces of
holonomies of the connection along loops. This technique had been used
in the context of Yang--Mills theories by Gambini and Trias
\cite{GaTr}.  The coefficients of the expansion of a wavefunction in
this basis are purely geometrical functions of the loops, that contain
the same information as the original wavefunctions. These coefficients
are the wavefunctions in the ``loop representation.'' This
representation is particularly attractive at the time of dealing with
another of the constraints of canonical gravity: the diffeomorphism
constraint. This constraint implies that the wavefunctions should be
functions of loops that are invariant under diffeomorphisms. Such
functions have been studied by mathematicians for some time, they are
called knot invariants. This implies a significant improvement over
the use of the traditional variables, since there the diffeomorphism
constraint did not encounter such a natural encoding in terms of the
wavefunctions.

Rovelli and Smolin also noted in 1995 \cite{RoSmspin} that one could
overcome a technical problem of the loop representation due to the
fact that the basis of loops is really an overcomplete basis (which
implies that certain relations exist between the coefficients in the
expansion).  They noted that if one used a mathematical construction
known as spin networks, introduced by Penrose in the 1960's, one could
label in a simple way the independent elements of the basis of
loops. Spin networks are diagrams made of lines that intersect at
vertices. Each line carries a holonomy in a certain representation of
the group and the lines are ``tied together'' at intersections using
invariant tensors in the group. Rovelli and Smolin \cite{RoSmarea}
noted that one could construct simple quantum operators associated to
the area of a surface and the volume of a region of space and these
operators, acting on spin network states had a simple action with
discrete eigenvalues.

The remaining constraint of canonical gravity is called the
Hamiltonian constraint. This constraint does not have a simple
geometrical interpretation and has to be implemented as a quantum
operator and studied. For years, this implementation remained elusive.
Most of the results in canonical gravity (with the notable exception
of the calculation of black hole entropies \cite{Ashtekar}) were
confined to ``kinematical'' results, i.e. results of considering
states that are not necessarily annihilated by the Hamiltonian
constraint.

Ashtekar and Lewandowski \cite{AsLe} were able to define a rigorous
integration theory in the space of functions of a connection modulo
gauge transformations, that acquires a particularly simple form when
presented in terms of spin networks. This allowed to rigorously show
that the spectrum of the area and volume are quantized \cite{AsLeavo}.

An implementation of the Hamiltonian constraint was finally found by
Thiemann in 1997 \cite{Th}. He noted certain classical identities
that allowed to write in an elegant way the singly-densitized
Hamiltonian constraint.  This operator is the only one that one can
naturally expect to have an action on spin network states where there
is no a priori defined background metric, since there the only 
naturally defined density is one of weight one: the Dirac delta function.

Thiemann's result is remarkable in many ways. To begin with, it
provides the first non-trivial, consistent, finite and well defined
theory of quantum gravity. Moreover, the action of the Hamiltonian
constraint on a spin network state is relatively simple: it acts at
vertices by adding a line connecting two of the incoming lines (it
produces a term for every pair of incoming lines) and multiplying
times a factor that depends on the valences of the lines. The big open
question at the moment is if Thiemann's Hamiltonian really captures
the correct dynamics of general relativity. It could be the case that
it provides a well defined theory but that it does not correspond to
a desirable quantization of general relativity.

There are some remarkable achievements already of Thiemann's theory.
In particular, Bojowald particularized the theory to homogeneous
cosmologies (unlike usual quantum cosmology in which one imposes
homogeneity classically and later quantizes, Bojowald first quantizes
and then restricts to states that represent homogeneous cosmologies,
this construction has been formalized recently
\cite{ashtekarbojowald}), and has found very attractive results. But
there are also some worries about Thiemann's theory. In particular, the
action of the Hamiltonian appears prima facie to be too simple. Since
it adds a line connecting two other lines, it generates new vertices
in the spin network that are planar. In particular this implies that
if one acts on a bra state with the operator, it removes a line
connecting two planar vertices. Now if one considers a bra state based
on a spin network of arbitrary complexity, but lacking planar
vertices, the Hamiltonian annihilates it. This seems to imply that
there are too many solutions.  This was observed by Thiemann as well
in $2+1$ dimensions \cite{thiemann21}. There the extra states can be
eliminated by a careful choice of inner product as not
normalizable. It is questionable if a similar construction will be
feasible in $3+1$ dimensions.

To try to understand Thiemann's theory, it is worthwhile examining the
spirit of the construction a bit. In order to regularize the Hamiltonian
constraint, Thiemann first discretizes the theory on a lattice and then
quantizes the theory. It is therefore of interest to analyze what kind of
theory is one actually quantizing on the lattice. This led us to consider
the subject of consistent discretizations.

\section{Consistent discretizations}

Discretizations are very commonly used as a tool to treat field theories.
Classically, when one wishes to solve the equations of a theory on a 
computer, one replaces the continuum equations by discrete
approximations to be solved numerically. At the level of quantization,
lattices have been used to regularize the infinities that plague field
theories. This has been a very successful approach for treating  
Yang--Mills theories.

Discretizing general relativity is more subtle than what one initially
thinks. Consider a $3+1$ decomposition of the Einstein equations. One
has twelve variables to solve for (the six components of the spatial
metric and the six components of the extrinsic curvature). Yet, there
are {\em sixteen} equations to be solved, six evolution equations for
the metric, six for the extrinsic curvature and four constraints. In
the continuum, we know that these sixteen equations are {\em
compatible}, i.e. one can find twelve functions that satisfy them.
However, when one discretizes the equations, the resulting system of
algebraic equations is in general incompatible. This is well known,
for instance, in numerical relativity. The usual attitude there is to
ignore the constraints and solve the twelve evolution equations (this
scheme is called ``free evolution''). The expectation is that in the
limit in which the lattice is infinitely refined, the constraints will
also be satisfied if one satisfied them initially. The situation is
more involved if one is interested in discretizing the theory in order
to quantize it.  There, one needs to take into account all
equations. In particular, in the continuum the constraints form an
algebra. If one discretizes the theory the discrete version of the
constraints will in many instances fail to close an algebra. Theories
with constraints that do not form algebras imply the existence of more
constraints which usually makes them inconsistent. For instance, it
might be the case that there are no wavefunctions that can be
annihilated simultaneously by all constraints. One can ask the
question if this is not happening in the construction that Thiemann
works out. To our knowledge, this issue has not been probed. What is
clear, is that discretizing relativity in order to quantize it will
require some further thinking.

The new proposal we have put forward \cite{GaPuprl}, called {\em
consistent discretization} is that, in order to make the discrete
equations consistent, the lapse and the shift need to be considered as
some of the variables to be solved for.  Then one has 16 equations and
16 unknowns. This might appear surprising since our intuition from the
continuum is that the lapse and the shift are freely specifiable. But
we need to acknowledge that the discrete theory {\em is a different
theory}, which may approximate the continuum theory in some
circumstances, but nevertheless is different and may have important
differences even at the conceptual level. This is true of any discretization
proposal, not only ours.

We have constructed a canonical approach for theories discretized in
the consistent scheme \cite{DiGaPu}. The basic idea is that one does
not construct a Legendre transform and a Hamiltonian starting from the
discretized Lagrangian picture. The reason for this is that the
Hamiltonian is a generator of infinitesimal time evolutions, and in a
discrete theory, there is no concept of infinitesimal. What plays the
role of a Hamiltonian is a canonical transformation that implements
the finite time evolution from discrete instant $n$ to $n+1$. The
canonical transformation is generated by the Lagrangian viewed as a
type I canonical transformation generating functional. The theory is
then quantized by implementing the canonical transformation as a
unitary evolution operator.

\section{Examples}

We have applied this discretization scheme to BF theory and
Yang--Mills theories \cite{DiGaPu}. In the case of BF theories this
provides the first direct discretization scheme on a lattice that is known
for such theories. In the case of Yang--Mills theories it reproduces
known results. We have also studied the application of the
discretization scheme in simple cosmological models. We find that the
discretized models approximate general relativity well and avoid the
singularity \cite{cosmo}. More interestingly, they may provide a
mechanism for explaining the value of fundamental constants
\cite{gapusmolin}. When the discrete models tunnel through the
singularity, the value of the lapse gets modified and therefore the
``lattice spacing'' before and after is different. Since in lattice
gauge theories the spacing is related to the ``dressed'' values of the
fundamental constants, this provides a mechanism for fundamental
constants to change when tunneling through a singularity, as required
in Smolin's \cite{smolin} ``life of the cosmos'' scenario.

It is quite remarkable that the discrete models work at all. When one
solves for the lapse and the shift one is solving non-linear coupled
algebraic equations. It could have happened that the solutions were
complex. It could have happened that there were many possible
``branches'' of solutions. It could have happened that the lapse
turned negative. Although all these situations are possible given
certain choices of initial data, it is remarkable that it appears that
one can choose initial data for which pathologies are avoided and the
discrete theory approximates the continuum theory in a controlled
fashion. 

We are currently exploring the Gowdy models with this approach. Here
the problem is considerably more complex than in cosmological
models. The equations to be solved for the lapse and the shift become
a coupled system that couples all points in the spatial discretization
of the lattice. The problem can only be treated numerically. We have
written a fortran code to solve the system using iterative techniques
(considerable care needs to be exercise since the system becomes
almost singular at certain points in phase space) and results are
encouraging. In the end the credibility of the whole approach will
hinge upon us producing several examples of situations of interest
where the discrete theories approximate continuum GR well.

\section{Several conceptual advantages}

The fact that in the consistent discrete theories one solves the
constraints to determine the value of the Lagrange multipliers has
rather remarkable implications. The presence of the constraints is one
of the most significant sources of conceptual problems in canonical
quantum gravity. The fact that we approximate the continuum theory
(which has constraints) with a discrete theory that is constraint free
allows us to bypass in the discrete theory many of the conceptual
problems of canonical quantum gravity. One of the main problems we can
deal with is the ``problem of time''. This problem has generated a
large amount of controversy and has several aspects to it. We cannot
cover everything here, the definitive treaty on the subject is the
paper by Kucha\v{r} \cite{Kuchar}.

To simplify the discussion of the problem of time, let us consider an
aspect of quantum mechanics that most people find unsatisfactory
perhaps from the first time they encounter the theory as
undergraduates. It is the fact that in the Schr\"odinger equation, the
variables ``$x$'' and ``$t$'' play very different roles. The variable
$x$ is a quantum operator of which we can, for instance, compute its
expectation value, or its uncertainty. In contrast $t$ is assumed to
be a continuous external parameter. One is expected to have a clock
that behaves perfectly classically and is completely external to the
system under study. Of course, such a construction can only be an
approximation. There is no such thing as a perfect classical clock and
in many circumstances (for instance quantum cosmology) there is no
``external clock'' to the system of interest. How is one to do quantum
mechanics in such circumstances? The answer is: ``relationally''. One
could envision promoting {\em all} variables of a system to quantum
operators, and choosing one of them to play the role of a
``clock''. Say we call such variable $t$ (it could be, for instance
the angular position of the hands of a real clock, or it could be
something else). One could then compute conditional probabilities for
other variables to take certain values $x_0$ when the ``clock''
variable takes the value $t_0$. If the variable we chose as our
``clock'' does correspond to a variable that is behaving classically
as a clock, then the conditional probabilities will approximate well
the probabilities computed in the ordinary Schr\"odinger theory. If
one picked a ``crazy time'' then the conditional probabilities are
still well defined, but they don't approximate any Schr\"odinger
theory well. If there is no variable that can be considered a good
classical clock, Schr\"odinger's quantum mechanics does
not make sense and the relational quantum mechanics is therefore
a generalization of Schr\"odinger's quantum mechanics.

Relational quantum mechanics therefore appears well suited as a
technique to use in quantizing general relativity, particularly in
cosmological situations where there is no externally defined
``classical time''. Page and Wotters advocated this in the 1980's
\cite{PaWo}.  Unfortunately, there are technical problems when one
attempts the construction in detail for general relativity. The
problem arises when one wishes to promote the variables to quantum
operators. Which variables to choose? In principle, the only variables
that make sense physically are those that have vanishing Poisson
brackets (or quantum mechanically vanishing commutators) with the
constraints. But since the Hamiltonian is one of the constraints, then
such variables are ``perennials'' i.e.  constants of motion, and one
cannot reasonably expect any of them to play the role of a
``clock''. One could avoid this problem by considering variables that
do not have vanishing Poisson brackets with the constraints. But this
causes problems. Quantum mechanically one wishes to consider quantum
states that are annihilated by the constraints. Variables that do not
commute with the constraints as quantum operators map out of the space
of states that solve the constraints. The end result of this, as
discussed in detail by Kucha\v{r} \cite{Kuchar} is that the
propagators constructed with the relational approach do not propagate.

Notice that all the problems are due to the presence of the
constraints. In our discrete theory, since there are no constraints,
there is no obstruction to constructing the relational picture. We
have discussed this in detail in \cite{greece}. 

Of great interest is the fact that the resulting relational theory
will never entirely coincide with a Schr\"odinger picture. In
particular, since no clock is perfectly classical, pure states do not
remain pure forever in this quantization, but slowly decohere into
mixed states. We have estimated the magnitude of this effect, and it
is proportional to $\omega^2 T_{\rm Planck} T$ where $
\omega$ is the frequency associated with the spread in energy 
levels of the system under study, $T_{\rm Planck}$ is Planck's time
and $T$ is the time that the system lives. The effect is very
small. Only for systems that have rather large energy spreads (Bose
Einstein condensates are a possible example) the effect may be close
to observability. With current technologies, the condensates do not
have enough atoms to achieve the energy spreads of interest, but it
might not be unfeasible as technology improves to observe the effect
\cite{deco}.

The fact that a pure state evolves into a mixed state opens other
interesting possibilities, connected with the black hole information
puzzle. This puzzle is related to the fact that one could consider a
pure quantum state that collapses into a black hole. The latter will
start evaporating due to Hawking radiation until eventually it
disappears. What one is left with at the end of the day appears to be
the outgoing radiation, which is in a mixed state. Therefore a pure
state appears to have evolved into a mixed state. There is a vast
literature discussing this issue (see for instance \cite{GiTh} for a
short review). Possible solutions proposed include that the black hole
may not disappear entirely or that some mechanism may allow pure
states to evolve into a mixed state. But we have just discussed that
the relational discrete quantum gravity predicts such decoherence! We
have estimated that the decoherence is fast enough to turn the pure
state into a mixed one before the black hole can evaporate completely,
at least if one considers black holes larger than a few hundred Planck
masses (for smaller holes the evaporation picture is not accurate
anyway) \cite{infopuzzle}. The result is quite remarkable, since the
decoherence effect, as we pointed before, is quite small. It is large
enough to avoid the information puzzle in black holes, even if one
considers smaller and smaller black holes which evaporate faster since
they also have larger energy spreads and therefore the decoherence
effect operates faster.

\section{Summary}

Analyzing the problems of the dynamics of loop quantum gravity led us
to develop a way to consistently discretize general relativity.
Surprisingly, the consistent discretizations not only approximate
general relativity well in several situations, but allow to handle
several of the hard conceptual problems of canonical quantum gravity.
What is now needed is to demonstrate that the range of situations in
which the discrete theory approximates general relativity well is
convincingly large to consider its quantization a route for the
quantization of general relativity.

\section{Acknowledgments}

This work was
supported by grant NSF-PHY0244335 and funds from the Horace Hearne
Jr. Institute for Theoretical Physics.

\end{document}